\documentclass{caosp302}
\usepackage{graphicx}
\firstpage{1}
\begin{document}

\hauthor{J.\,Braithwaite}
\title{Non-dipolar magnetic fields in Ap stars}
\author{J.\,Braithwaite \inst{1}}
\institute{Canadian Institute for Theoretical Astrophysics\\
  60 St.~George St., Toronto ON M5S 3H8, Canada\\
  \email{jon@cita.utoronto.ca}}
\date{January 29, 2008}
\maketitle

\begin{abstract}
An arbitrary initial magnetic field in an A star evolves into a stable equilibrium. Simulations are presented of the formation of non-axisymmetric equilibria consisting of twisted flux tubes meandering under the surface of the star, and analytic arguments are given relating the stability and form of these equilibria. These results may help to explain observations of Ap stars with very non-dipolar fields. This work is also applicable to other essentially non-convective stars such as white dwarfs and neutron stars.
\end{abstract}

\section{Introduction}

Convective stars of various types tend to have small-scale, time-varying magnetic fields, which is interpreted as dynamo action driven by differential rotation and convection/buoyancy instabilities. In contrast, intermediate-mass main-sequence stars ($1.5M_\odot<M<6M_\odot$), which have a small convective core and a radiative envelope, have either no detectable field or a large-scale, steady magnetic fields -- the Ap stars (see e.g. Auri\`ere et al. 2008). Many of these stars have roughly dipolar fields but many have rather more complex fields, such as $\tau$ Sco (Donati et al. 2006).

Given the lack of necessary dynamo ingredients, we infer that the magnetic fields are `fossil' remnants in some stable equilibrium, formed from the magnetic field left over from the convective/accretive protostellar phase. On the theoretical side, the emphasis has been on finding stable magnetic equilibria. Analytically one can produce an equilibrium configuration simply by making sure all forces are balanced, and then test its stability using an energy method (Bernstein et al. 1958). Unfortunately, it has been easier to demonstrate the {\it instability} of various equilibria than to find {\it stable} configurations. For instance, both axisymmetric poloidal fields and axisymmetric toroidal fields have been shown to be unstable (Wright 1973; Tayler 1973; Braithwaite 2006). More recently, it has become possible to find stable equilibria using numerical methods. Braithwaite \& Nordlund (2006, see also Braithwaite \& Spruit 2004) found such a configuration by evolving an arbitrary initial magnetic field in time and watching it find its way into a stable equilibrium. This equilibrium is roughly axisymmetric, consisting of both toroidal and poloidal components (logical, since both are unstable on their own) in a twisted-torus configuration. From the outside, the field looks approximately dipolar, as do the fields of many Ap stars. Here, I demonstrate the existence of more complex, non-dipolar configurations. I first describe simulations of magnetic equilibria evolving from arbitrary initial conditions, then use simple analytic methods to explore the nature and stability of these equilibria.

\section{Numerical simulations}

The setup of the simulations is described in detail in Braithwaite \& Nordlund (2006); I give a brief summary here. The code used is the {\sc stagger code} (Gudiksen \& Nordlund 2005), a high-order finite-difference Cartesian MHD code. The star is modelled as a ball of self-gravitating ideal gas of radius $R$ arranged in a polytrope of index $n=3$ (approximating an A star), surrounded by an atmosphere of very low electrical conductivity. The initial magnetic field resembles that expected at the end of a period of convection, i.e. chaotic and small-scale, containing wavenumbers with a flat power spectrum up to $k_{\rm max}$ (set to $36R^{-1}$ here), tapered so that $B \sim \rho^p$ where $\rho$ is the gas density and $p$ is a free parameter. Note that if the star forms out of a uniformly magnetised cloud and the same fraction of flux is lost from all fluid elements, then we expect $p=2/3$.

The evolution in time of the magnetic field is followed in simulations with various values of the tapering parameter $p$: $0$, $1/3$, $2/3$ and $1$. In all cases, a stable equilibrium is reached after a few Alfv\'en crossing times, but both the geometry and energy of the equilibrium is found to vary. Two types of equilibrium are found: simple axisymmetric (as in Braithwaite \& Nordlund 2006) in the $p=1$ and $p=2/3$ cases, and more complex non-axisymmetric in the $p=1/3$ and $p=0$ cases.

\begin{figure}
\includegraphics[width=0.45\hsize,angle=0]{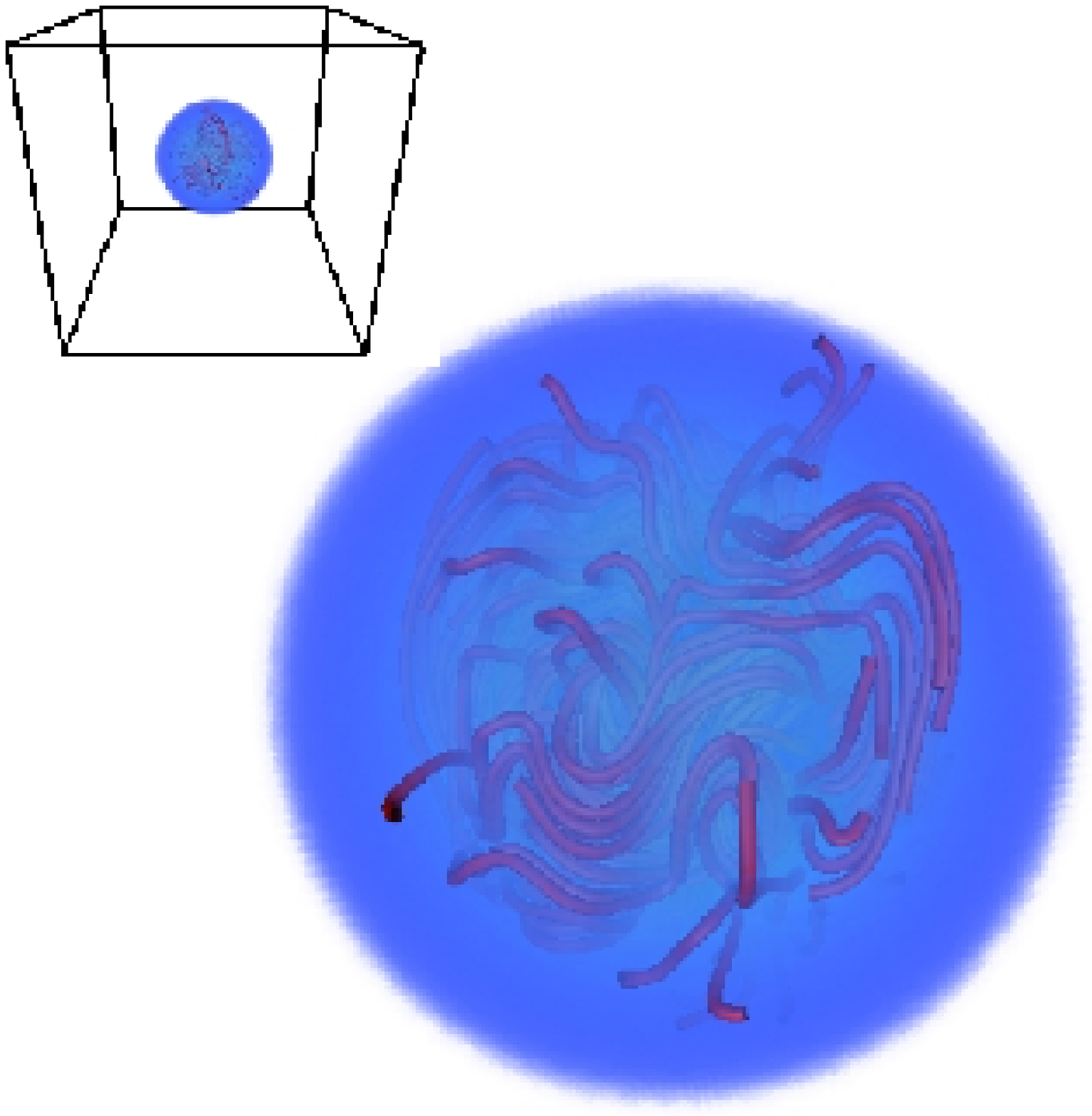}\hfill
\includegraphics[width=0.45\hsize,angle=0]{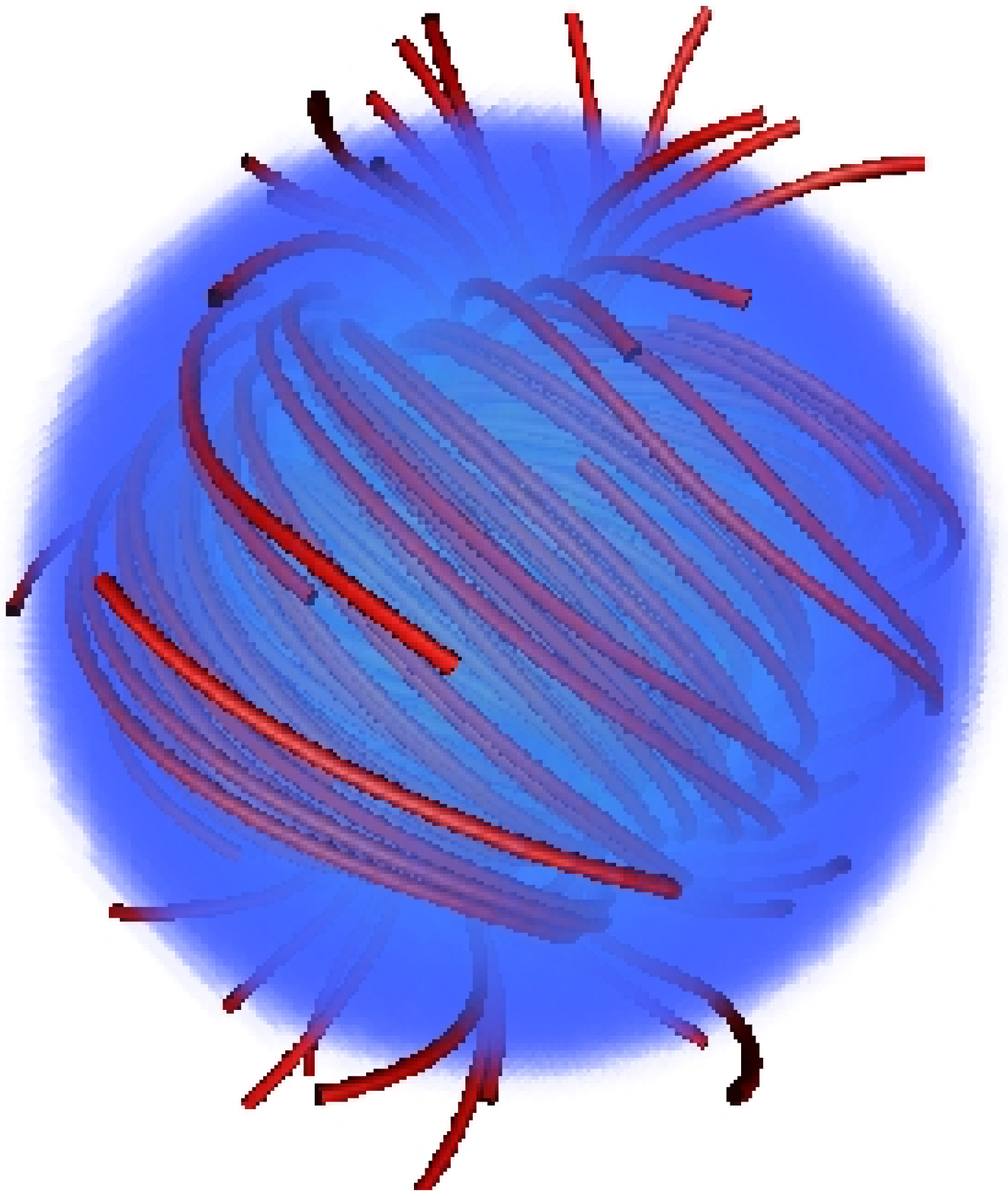}
\includegraphics[width=0.45\hsize,angle=0]{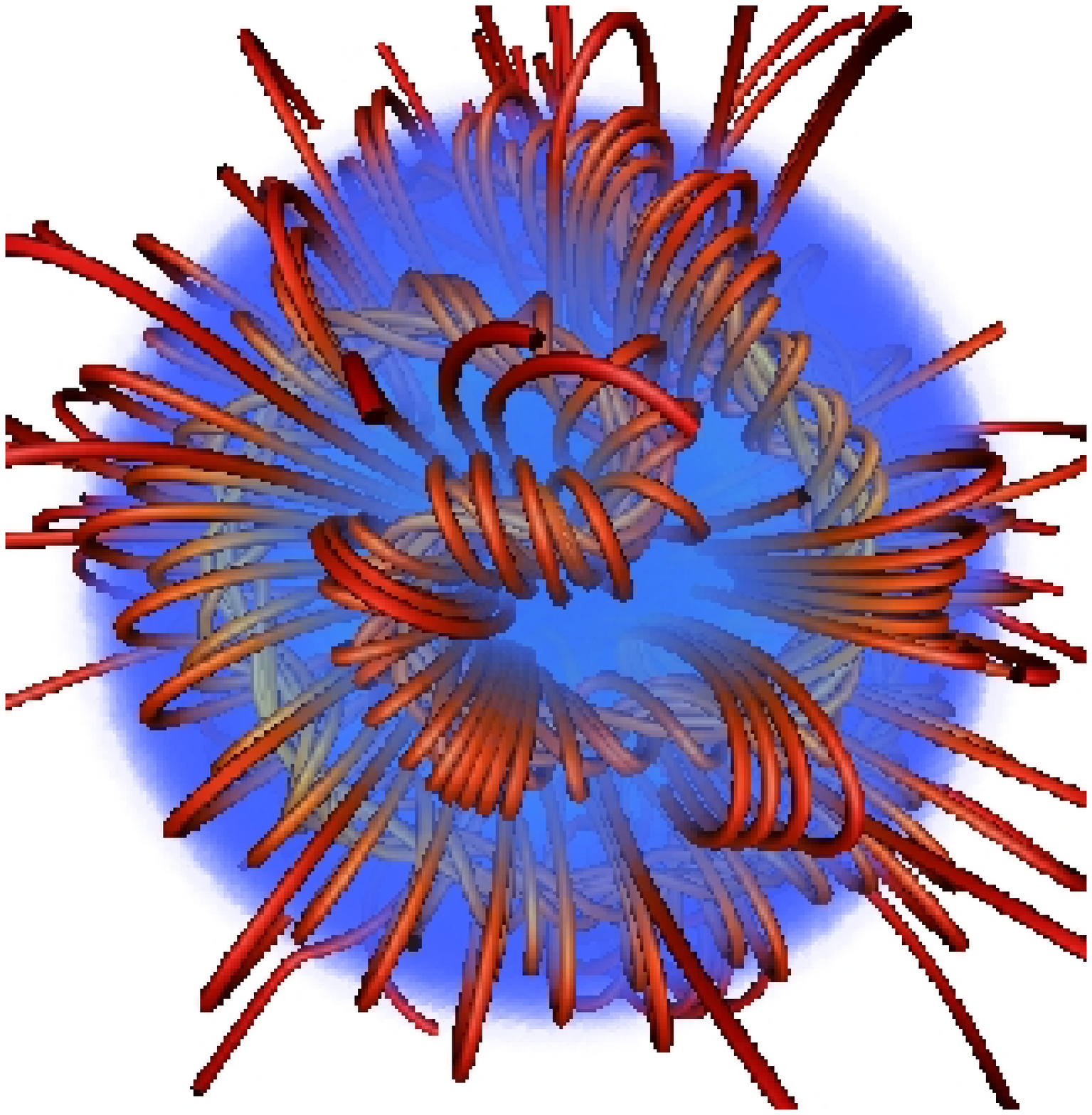}\hfill
\includegraphics[width=0.45\hsize,angle=0]{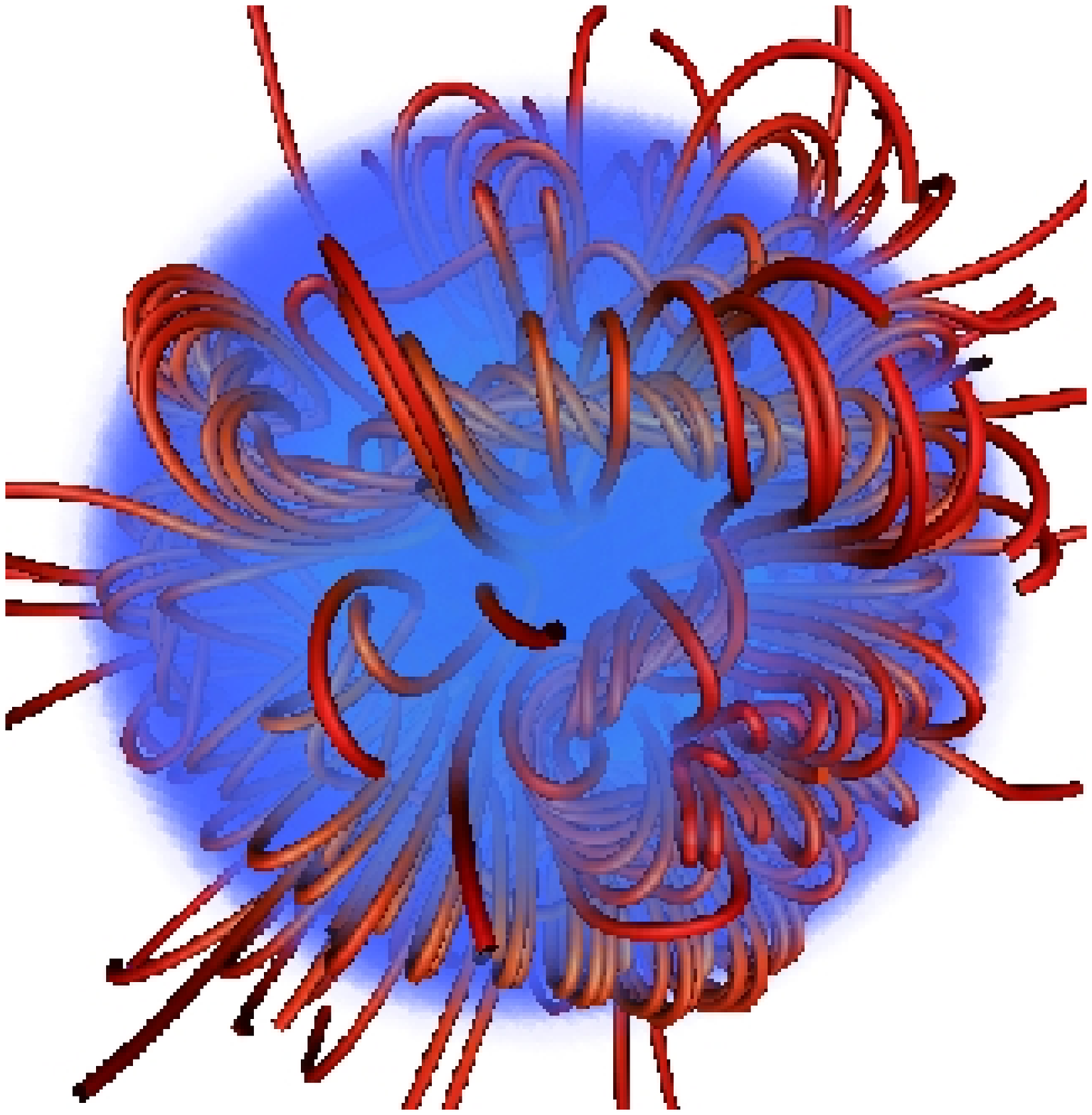}
\caption{Top-left, the initial conditions. Semi-opaque shading represents the star; note that there is a lot of free space around the star, which ensures an accurate treatment of the potential (current-free) field outside the star. Top-right, the equilibrium reached in the case of a centrally-concentrated ($p=1$) initial field. Bottom left and right (viewed from either side of the star) is the equilibrium reached in the $p=0$ case.}
\label{fig:vapor}
\end{figure}

Fig.~\ref{fig:vapor} shows the initial conditions (top-left), as well as both the $p=1$ (top-right) and $p=0$ (bottom-left \& -right) runs after around $9$ Alfv\'en crossing times. In the $p=1$ case, the configuration is roughly axisymmetric and contains both toroidal and poloidal components. When $p=0$, a non-axisymmetric equilibrium is reached. The configuration consists of twisted flux tubes below the surface of the star. The flux tubes lie horizontally and meander around the star at some distance below its surface. 

\section{Stability of non-axisymmetric fields}

\begin{figure}
\includegraphics[width=1.0\hsize,angle=0]{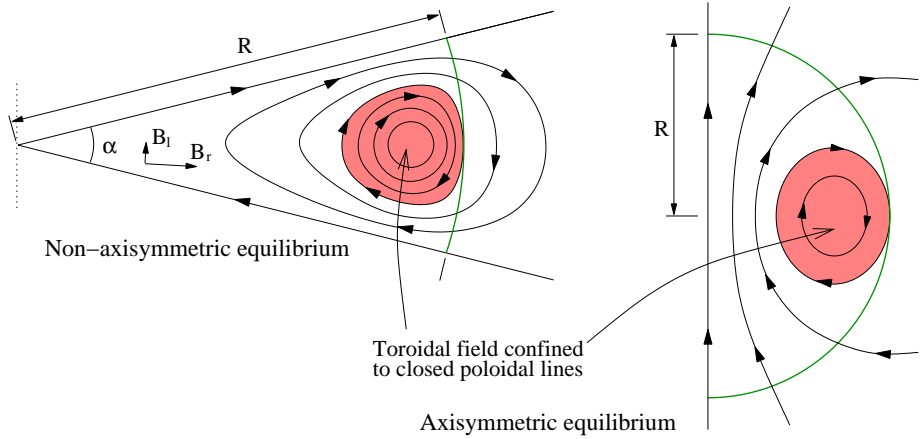}
\caption{Cross-section of a flux tube in a star. In the axisymmetric case, the tube is a circle around the equator; in the non-axisymmetric case, the tube meanders around the star in a more complex fashion. Directions of $B_{\rm l}$ and $B_r$ are indicated by arrows towards the left; $B_{\rm t}$ is directed into the page. Poloidal field lines are marked with arrows; toroidal field is represented by shading.}
\label{fig:x-sec}
\end{figure}

Consider a twisted flux tube of length $s$ and width (at the surface of the star) $\alpha R$ whose axis lies at some constant radius in a star of radius $R$. The cross-section of such a flux tube is illustrated in Figure~\ref{fig:x-sec}, as well as its axisymmetric relative. It is assumed that $\alpha \ll1$. The three components of the magnetic field are toroidal (parallel to the length of the flux tube), latitudinal and radial; average values are denoted respectively by $B_{\rm t}$, $B_{\rm l}$ and $B_r$. The latitudinal and radial components are collectively referred to as the poloidal component. Since the flux tube cannot have any ends, it must be joined in a loop, which can be wrapped around the star in any fashion.

It is shown in Fig.~\ref{fig:x-sec} that the toroidal component of the field is confined to the volume enclosed by the largest poloidal line closed within the star, and is zero outside this volume. This has to be the case, because a poloidal line which crosses the stellar surface would `unwind' if it contained any toroidal component. In this process, toroidal field is redistributed along the whole of the poloidal line. Toroidal field in the atmosphere is then destroyed by reconnection on a short time-scale, and the toroidal field inside the star is redistributed along the poloidal line to where it has been lost. This occurs on the Alfv\'en time-scale. In this way, all toroidal field on the poloidal line is quickly transferred into and destroyed in the atmosphere.

The magnetic energy of the flux tube is
\begin{equation}
E = \int{\frac{B^2}{8 \pi} dV} \approx \frac{s \alpha R^2}{3} \frac{1}{8 \pi} \left[ B^2_{\rm t} + B^2_{\rm l} + B^2_r \right],
\label{eq:energy}
\end{equation}
where $s$ is the length of the tube (on the surface of the star), $\alpha$ is the angle subtended at the centre of the star. The three terms on the right hand side represent the field energy in the three components. The field energy outside the star can be ignored (Braithwaite 2008, in prep.).

Allowing the length and width of the flux tube to change adiabatically, the tube will stretch or contract until $\partial E/\partial \alpha=0$, i.e. an equilibrium is reached\footnote{It is straightforward to verify that $\partial^2 E/\partial \alpha^2 > 0$ at this point, i.e. that the equilibrium is stable to stretching/contracting perturbations.}. Since we are considering a magnetic field in a stably-stratified star where the magnetic energy density is very much less than the thermal, i.e. $B^2/8\pi \ll P$, we have the following restrictions on the displacements $\mathbf{\xi}$ which take place during this adjustment:
\begin{equation}
{\bf \nabla}\cdot{\bf \xi} \approx 0 \;\;\;{\rm and}\;\;\; {\bf g}\cdot{\bf \xi} \approx 0.
\end{equation}
The consequence of these restrictions is that the volume of the flux tube stays constant since no matter can leave or enter it, which can be expressed by stating that the area of the flux tube on the stellar surface $F = s \alpha R = \mathrm{const}$. In addition, we have:
\begin{equation}
\frac{\partial \ln B_{\rm t}}{\partial \ln \alpha} = -1 \:,\;\;\;\;\frac{\partial \ln B_{\rm l}}{\partial \ln \alpha} = 1 \;\;\;\;{\rm and}\;\;\;\; \frac{\partial \ln B_r}{\partial \ln \alpha} =  0,
\label{eq:ddalpha}
\end{equation}
which follow simply from flux-freezing. Differentiating (\ref{eq:energy}) with respect to $\alpha$ and setting $\partial E/\partial \alpha=0$, and noting from the geometry that $B_{\mathrm l} \approx \alpha B_r$, we have
\begin{equation}
B_{\mathrm t} \approx B_{\mathrm l} \approx \alpha B_r.
\label{eq:BtBl}
\end{equation}
Therefore the adjustment to equilibrium consists in stretching or contracting until the toroidal and latitudinal components are roughly equal.

The toroidal and poloidal fluxes in the tube are $\Phi_{\rm t} \approx (1/2)\alpha R^2 B_{\rm t}$ and $\Phi_{\rm p} \approx (1/2)s\alpha R B_r = (1/2) F B_r$ respectively. They are conserved during adiabatic changes, and using (\ref{eq:BtBl}) we have
\begin{equation}
\alpha^2 \approx \frac{2}{R^2} \frac{\Phi_{\rm t}}{B_r} \approx \frac{F}{R^2} \frac{\Phi_{\rm t}}{\Phi_{\rm p}},
\label{eq:alpha2}
\end{equation}
so that if we know the area $F$ and the ratio of toroidal to poloidal fluxes of a tube, we can calculate its equilibrium width $\alpha R$ and length $s$.

The above arguments can be applied not only to the flux tube as a whole, but also to any small length $\delta s$ of the tube with area $\delta F$. Along the length of the flux tube $\Phi_{\rm t}$ is constant, but $B_r \approx 2 \delta \Phi_{\rm p}/\delta F$ can vary. A single flux tube need not have uniform width -- at every point along its length, there is a local equilibrium giving $\alpha$ in terms of the toroidal flux and the local value of $B_r$ (remember from (\ref{eq:ddalpha}) that $B_r$ is constant during adiabatic changes in $\alpha$).

In a star with more than one flux tube, neighbouring flux tubes must have their poloidal components in opposite senses to avoid boundary discontinuities. However, there is no constraint on the sense of the toroidal component, since the tubes are separated from each other by a region with zero toroidal component. Since magnetic helicity is essentially just the product of the poloidal and toroidal components, different tubes can have helicities of opposite sign, and configurations of zero net helicity are possible in principle. This contradicts the hypothesis that an arbitrary field should evolve into the lowest energy state for its helicity, rather the field can find local energy minima at higher energies.

\section{Discussion and conclusions}

Once an equilibrium has formed, it continues to evolve as a result of magnetic and thermal diffusivities ($\eta$ and $\kappa$ respectively). Magnetic diffusion (a result of finite conductivity) causes the field to evolve on a timescale $\mathcal{L}^2/\eta$ where $\mathcal{L}$ is the relevant length scale. For an A star, this is around $10^{10}$yr if $\mathcal{L}=R$. Thermal diffusion can also cause field evolution. To be in pressure and density equilibrium with its non-magnetised surroundings, a flux tube must be at a lower temperature than its surroundings. Heat diffuses into the tube, causing it to rise on a timescale of order $\mathcal{L}^2\beta/\kappa$, where $\beta$ is the ratio of thermal to magnetic pressure. These effects are discussed in Braithwaite 2008 (in prep.); it is likely that both are too slow to be important in A stars, except perhaps in stars with very complex, small-scale equilibria (i.e. with smaller $\mathcal{L}$). Whether there is any observational evidence for field evolution in A stars is controversial.

That we observe both axisymmetric and non-axisymmetric fields in Ap stars implies that initial conditions vary, and that probably the central concentration of the initial field varies. The reason for this is unclear. However, it is a common phenomenon that Ap stars share with white dwarfs and neutron stars (as is the enormous range in field strengths), while the apparent cutoff in field strengths below $300$G (Auri\`ere et al. 2008) is unique to A stars, so the two problems are probably unconnected.

In summary, I have performed numerical simulations of the formation of stable magnetic equilibria in a non-convective star, starting from turbulent initial conditions. A sufficiently centrally-concentrated initial field evolves into an approximately axisymmetric equilibrium; a more spread-out initial field evolves into a more complex, non-axisymmetric equilibrium consisting of twisted flux tubes meandering under the surface of the star. In this case, the adjustment to equilibrium consists in the stretching or contracting of these flux tubes until an energy minimum is reached. Still needed is a more quantitative understanding of the effect of initial conditions on the resulting equilibrium, and more work is also required on the quasi-static diffusive evolution of these equilibria. The results are applicable to all predominantly non-convective stars: main sequence above $1.5M_\odot$, white dwarfs and neutron stars.

\acknowledgements
The author thanks Henk Spruit, Chris Thompson and Gregg Wade for fruitful discussions. Fig.~\ref{fig:vapor} was made with {\sc vapor} ({\tt www.vapor.ucar.edu}).

\end{document}